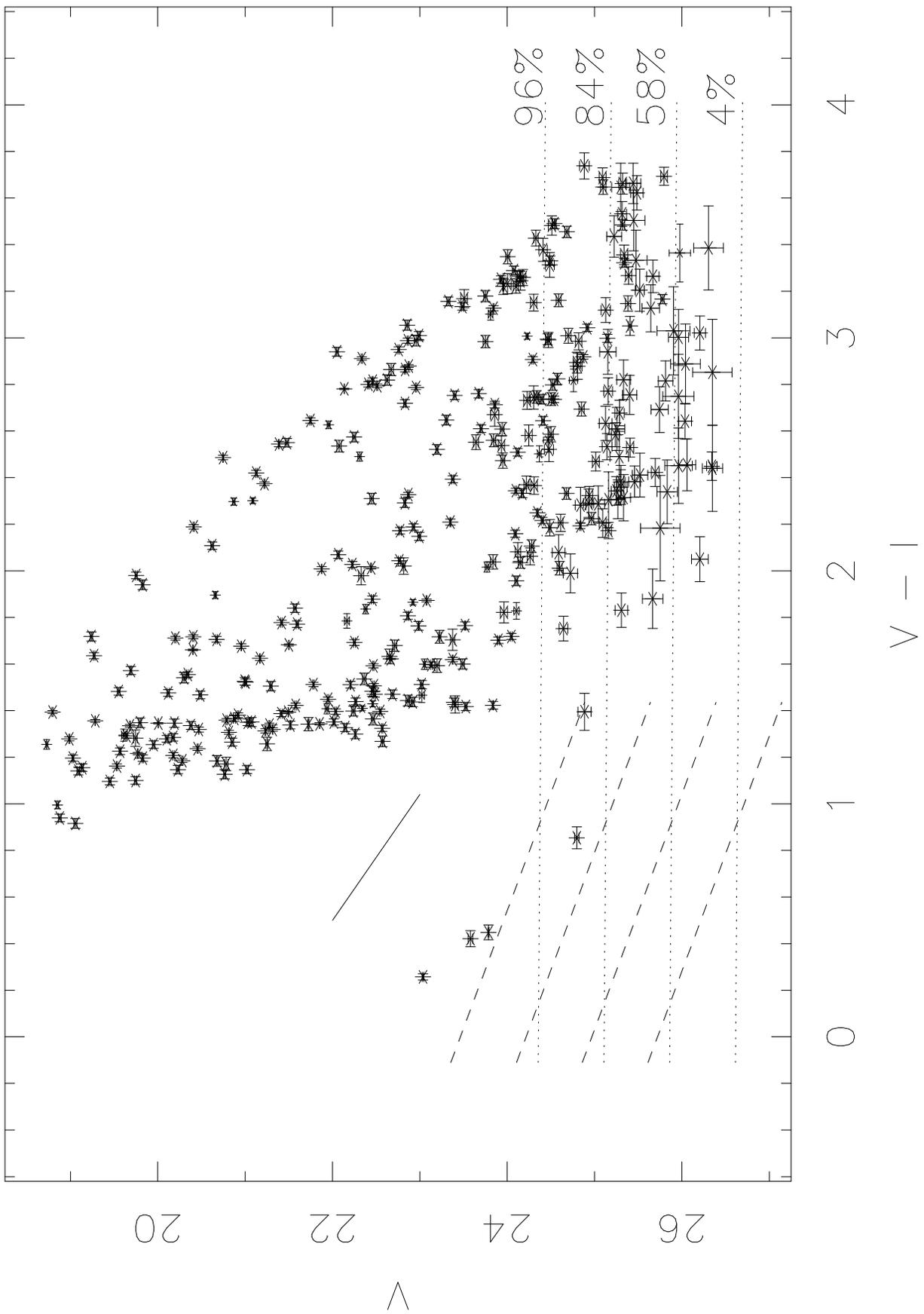

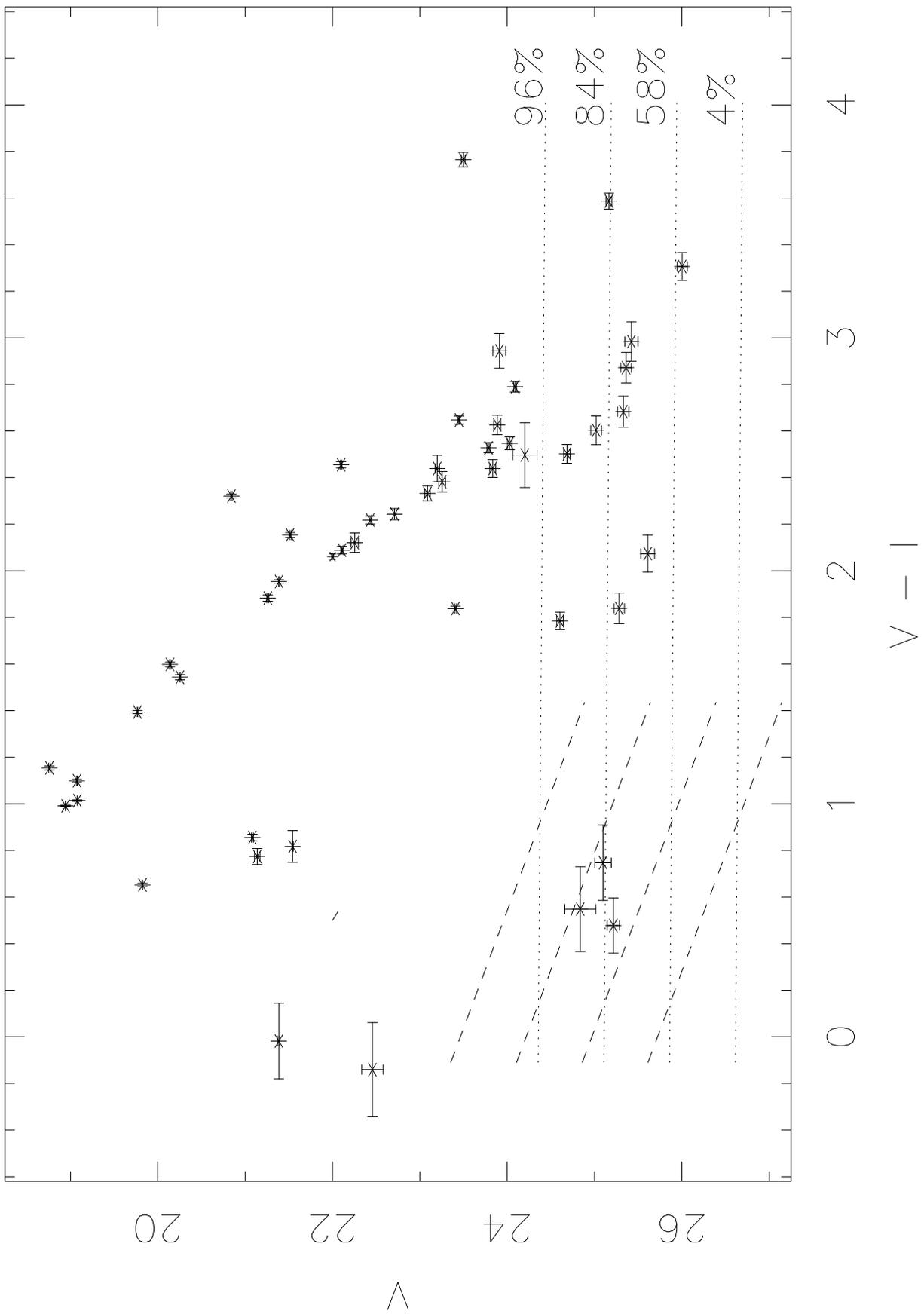

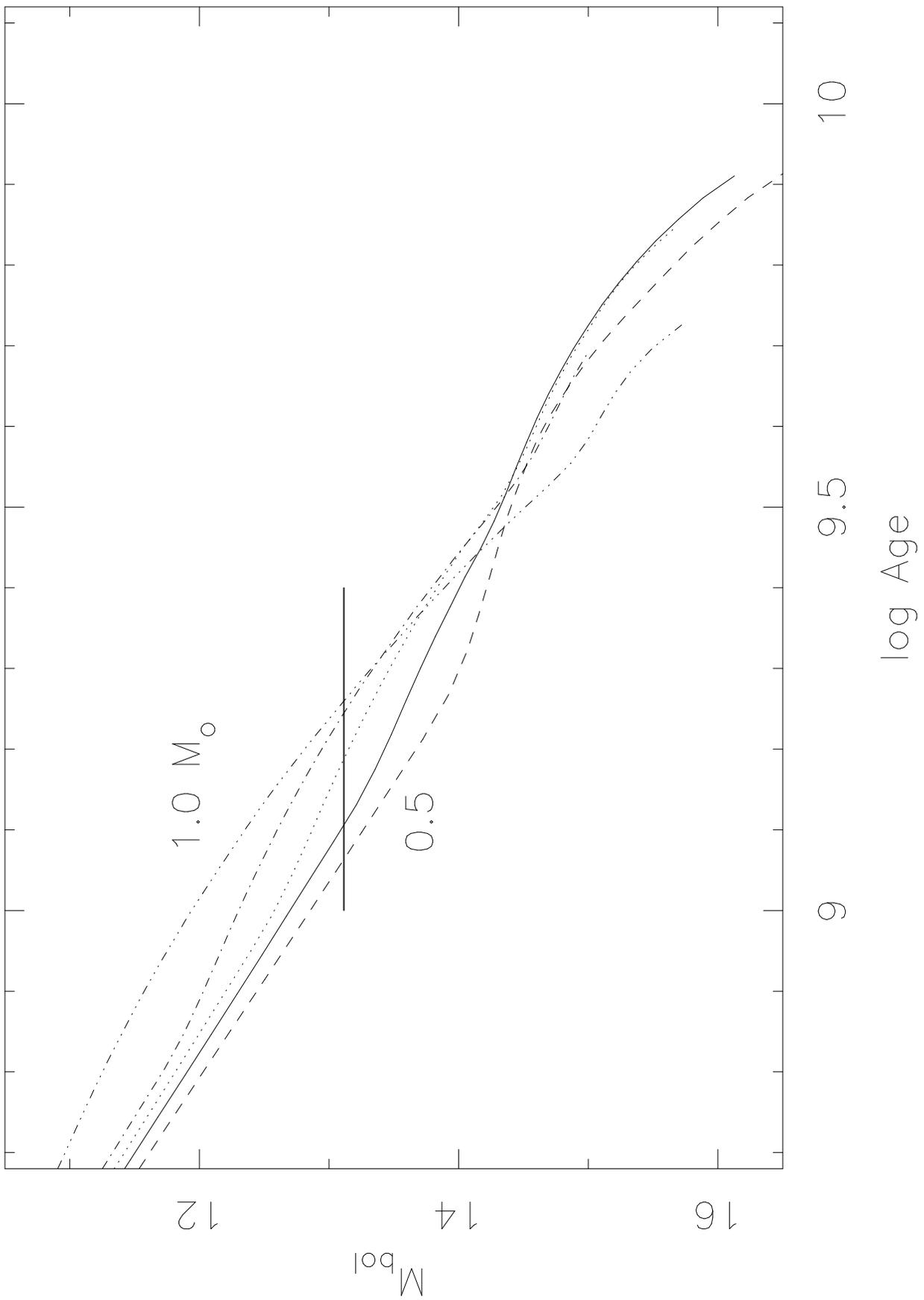

# An Independent Calibration of Stellar Ages: HST Observations of White Dwarfs at V=25

Ted von Hippel[1,2], Gerard Gilmore[1], D.H.P. Jones[1,3]

[1] *Institute of Astronomy, Madingley Road, Cambridge CB3 0HA*
[2] *WIYN Telescope, National Optical Astronomy Observatories, PO Box 26732, Tucson, AZ 85726-6732, USA*
[3] *Royal Greenwich Observatory, Madingley Road, Cambridge CB3 0EZ*

**ABSTRACT**

The white dwarf luminosity function of a stellar cluster will have a sharp truncation at a luminosity which is determined by the time since formation of the first white dwarfs in that cluster. Calculation of the dependence of this limiting luminosity on age requires relatively well-understood physics and is independent of stellar evolutionary models. Thus, measurement of the termination of the white dwarf luminosity function provides an independent method to determine the age of a cluster, and thereby to calibrate stellar evolutionary ages. We have obtained HST WFPC2 data in two open clusters, identified the white dwarf sequence, and proved the feasibility of this approach, by detecting white dwarfs to V=25. Much deeper data are feasible. From our present limited data, we show that degenerate cooling ages are not consistent with some published isochrone ages for clusters with ages of order 1Gyr.

## 1 INTRODUCTION

The application of stellar evolutionary models to observational data for open and globular clusters has reached a very considerable degree of sophistication and success. Continuing challenges for such models relate to their calibration – the precision with which the Sun and well-studied wide binaries can be described (*cf* Andersen 1991) as well as the relevant physical processes which need consideration only for some mass ranges – for example convective overshooting. The complexity of these problems is such that the several major groups producing isochrones for the determination of stellar ages are rarely in agreement at the 25 percent level, with occasionally even larger disagreements. The significance of this uncertainty is exacerbated by suggestions that the Hubble age is as much as a factor of two less than the isochrone ages for the oldest globular clusters. Before such a result can be considered robust, it is necessary to ensure that extant stellar age determinations are free of significant systematic errors. Thus, an independent calibration of isochrone ages is highly desirable.

Such a calibration is available, using white dwarf cooling tracks. Once a white dwarf is formed, there is a very simple relationship between its age and luminosity. This provides a clock which is able in principle to calibrate the isochrone ages. What is required is to determine the luminosity, and thereby the age, of the coolest white dwarfs in stellar clusters with a range of isochrone ages. Since the time taken for the stellar precursor of the first white dwarf to evolve is short, any uncertainty in this stellar age is an insignificant uncertainty in clusters of age greater than about a Gyr. Additionally, for intermediate age white dwarfs, the age-luminosity relationship is sufficiently steep that age determination with a formal precision of $\sim 0.2$Gyr is possible. Thus, one may use ages determined from the relatively simple physics of white dwarf cooling to calibrate age determinations from the complex physics of stellar evolution.

This experiment has been carried out for field stars in the Solar Neighbourhood (Winget *et al* 1987). The initial analysis suggested a white dwarf cooling age of $9.3\pm2.0$Gyr, which is some 2/3 that of the isochrone ages of the disk globular clusters. Later analyses (Yuan 1989) of the same data set preferred a broader age range, with $14 > \tau_{disk} > 6$Gyr. The uncertainty here arises from both white dwarf cooling theory, which is problematic at very low temperatures, where internal crystallisation is significant (eg Canal 1990), and from some sensitivity to the Galactic star formation history, and to the distribution of masses of white dwarfs (Yuan 1989, Weidemann 1990). In addition, small number statistics combine with the low intrinsic luminosity of the white dwarfs to make the experiment challenging.

The observational challenges are even more daunting in the nearest intermediate age open clusters, where the coolest white dwarfs will be seen near V=25. The real limit to precise age determinations is thus set by the precision with which one can identify the low luminosity limit of the white dwarf luminosity function. We present here the first results from observations of two open clusters with the Hubble Space Telescope which reach at or near the limit of the white dwarf cooling sequence, proving the viability of this new dating technique. HST is required as its high image quality allows both faint limiting magnitude and the spatial resolution necessary to remove the numerous background galaxies which contaminate the colour-magnitude diagram.

## 2   OBSERVATIONS AND DATA REDUCTION

We observed two open clusters, NGC 2477 and NGC 2420, with HST on 18 March 1994 and 18/19 May 1994, respectively, using the WFPC2 along with the F555W (approximately V-band) and F814W (approximately I-band) filters. Due to the large number of very bright stars in these clusters we obtained cumulative one hour exposures in each filter subdivided as nine 400 second exposures in NGC 2477 and four 900 second exposures in NGC 2420. In March the CCDs were operated at −77 C, while in May they were operated at −88 C. The temperature change was in response to a positionally-dependent non-linearity in flux measurements, apparently due to a charge-transfer problem. The maximum effect of this non-linearity was measured to be ≈10 percent with the CCDs at −77 C and ≈4 percent with the CCDs at −88 C. The cooled CCD also suffered far fewer "hot pixels" (pixels with an elevated dark count for hours or days) than the warmer CCD. An additional difference between our two clusters is that the sky levels for the May data were five times higher than for the March data, due to zodiacal light differences.

We reprocessed the raw data to include correction for the non-linearities in the March data, using the IRAF STSDAS package. The data reduction steps included the standard statistical corrections to the Analogue to Digital conversion, bias subtraction, dark count subtraction and flat fielding. We also included the non-standard "ramp correction" for the March charge-transfer problem. No correction is yet available for the May data. The effect on the May data is expected to be ≤1 percent because of the higher sky counts and the cooler CCDs. We additionally spent great effort trying to identify hot pixels as measured in the dark frames taken near our observations, but were largely unsuccessful due to their short time-dependence. Since hot pixels have counts which are independent of the filter in use, and thus have colour equal to zero on the instrumental system, they unfortunately appear in the same region of the colour-magnitude diagram as the faintest white dwarfs. Since the WFPC2 CCDs are undersampled, hot pixels may appear like faint stars, especially two adjacent hot pixels or single hot pixels next to positive sky fluctuations.

Aperture photometry was performed using 1.5 pixel apertures with aperture corrections based on a detailed study of the many bright stars available in our fields. We did not employ point spread function (psf)-fitting photometry after some experimenting as our fields were uncrowded and the psf was positionally dependent and not well enough characterized by our stars. The photometry was first calibrated to the WFPC1 ground system using the transformations given by Holtzman *et al* (1995) and then further converted to the Johnson-Cousins system using the transformations given by Harris *et al* (1993). The colour-magnitude diagrams for all the stars in the fields of both our clusters, with galaxies removed, converted to the WFPC1 ground system, are given in figures 1 and 2.

In order to estimate our completeness limits we added simulated stars created by the HST image generation software package TINYTIM at various flux levels and tried to recover them using our standard finding and photometry procedure. From this procedure we estimated our completeness levels, which are superimposed on figures 1 and 2. These completeness levels lead to a limiting magnitude ≈ 0.5 magnitudes brighter than initially expected for the refurbished HST, but this is now understood as due to CCDs leaking some charge between pixels (Holtzman *et al* 1994). Our estimates of completeness are probably accurate for the redder stars but not for the WDs. This is due to the hot pixels mentioned above which we estimate to be indistinguishable from WDs in our data just a few tenths of a magnitude below our observed WD limit. Our WD limits are thus upper estimates of the terminus of the cluster WD cooling sequences. Further details on the data reduction, and completeness calculations, together with an analysis of the cluster main sequence luminosity functions, can be found in von Hippel *et al* (1995).

## 3   CLUSTER ISOCHRONE AGES

Isochrones and corresponding ages for open clusters have been produced by several groups recently. Examples include Bertelli *et al* (1994), Chieffi & Straniero (1989), Demarque, Green & Guenther (1992), Maeder & Meynet (1992) and VandenBerg (1985). For present purposes the wealth of information in these isochrones provides a range of alternative ages for a given cluster.

NGC 2420 is a relatively young low abundance open cluster, at distance modulus 11.95. It is only lightly reddened, with a uniform $E_{B-V} = 0.05$, and has [Fe/H]≈ −0.45. Ages have been derived from good photometry by VandenBerg (1985), by Anthony Twarog *et al* (1990) and by Carraro & Chiosi (1994). VandenBerg, from earlier published photometry, derived an age of 4Gyr, and noted the good agreement between data and models. The Carraro & Chiosi result, from a set of isochrones which is generally consistent with the younger age scale of clusters, provides an age of 2.1Gyr. The Anthony Twarog *et al* analysis utilised the isochrones of VandenBerg (1985), but shifted, to fix them to a Solar calibration. This then led to an age of 3.4 ± 0.6Gyr, although they note a 'large uncertainty because of the inability of the standard isochrones to match the shape of the cluster turnoff'. This difference of 1.2Gyr in isochrone age from the more recent data corresponds to a difference in predicted white dwarf limiting luminosity, other things being equal, of 0.6magnitudes.

NGC 2477 has distance modulus about 10.6, [Fe/H]≈ 0, and uneven reddening, with an adopted $E_{B-V} = 0.33$ (Smith & Hesser 1983; Hartwick, Hesser & McClure 1972). Smith & Hesser derive an age of 1.2 ± 0.3Gyr, using the early Ciardullo & Demarque (1977) isochrones. Carraro & Chiosi (1994) using the Hartwick *et al* data, but using new isochrones including convective overshoot (Bertelli *et al* 1994), derive an age of 0.6±0.1Gyr. This factor of two difference in isochrone age corresponds to a difference in limiting white dwarf luminosity of approximately one magnitude.

The three predicted white dwarf limiting luminosities for NGC 2420, and the two predictions for NGC 2477 are summarised in Table 1, in the column labelled $M_{bol,\text{WD}}$

Figure 1: Colour Magnitude data for NGC 2477. The data are in the WFPC1 ground photometric system. The cluster reddening vector (solid line) and the data completeness limits for V (dots) and I (dashes) are shown. We place the white dwarf detection limit near V=25, V–I=0.9

(Predicted).

## 4   CLUSTER WHITE DWARF OBSERVATIONS

The CM data of figures 1 & 2 identify a (sparsely populated) white dwarf cooling sequence for each cluster. The limiting magnitude of this sequence may be derived, since the expected luminosity function is tolerably known, by maximum likelihood deconvolution. In the interim, pending larger area and deeper coverage, we proceed by deriving two preliminary estimates of the limiting flux which bracket the optimum estimator. An estimator which will provide a *bright* limit is found by simply determining the best-fit gaussian which is consistent with the data for the coolest few white dwarfs, allowing for the known observational errors. This is a biased estimator, in that the degenerate luminosity function is not considered. The luminosity function will be such that the numbers of degenerates per luminosity interval increase to a sharp cutoff. Thus, fitting a low order moment assuming an underlying delta function distribution will be biassed away from the true cutoff.

A second estimator, biased to provide a *faint* limit, is provided by the location of the coolest degenerate seen. Since the luminosity function is maximum at the cool limit, this star will statistically be most likely drawn from that limit. However, the one-sided bias in the underlying distribution function will bias the observed position of the coolest star to lie preferentially below the underlying peak luminosity.

In order to compare the data to white dwarf cooling models the bolometric magnitudes of the white dwarfs were computed by subtracting the distance modulus and using the absolute visual magnitude - absolute bolometric magnitude relation for white dwarfs given by Liebert, Dahn and Monet (1988).

The two estimators have been derived, together with their formal internal uncertainties, and are presented in table 1, in the columns labelled $M_{bol,\text{WD}}$ (Mean of faintest WDs) and $M_{bol,\text{WD}}$ (Coolest single WD).

## 5   DEGENERATE *VS* ISOCHRONE AGES

The rate of white dwarf cooling, the duration of stellar evolution prior to white dwarf formation, and the mass of the resulting white dwarf, are all functions of the initial mass of the star. One must then combine all three dependancies to determine the mass, and hence the cooling age, of the least luminous white dwarf in a cluster, and from that in turn the age since formation of the most massive progenitor whose degenerate descendant is observed. We adopt the white dwarf cooling curves of Wood (figure 3), the initial-mass – final-mass relationship determined empirically by Reimers & Koester (1982) and Weidemann & Koester (1983; see also Weidemann 1990), and for consistency with our adopted isochrone ages, the pre-white dwarf evolutionary lifetimes of Bertelli *et al* (1994). For the present data, so long as the initial mass of the most massive star which generates a white dwarf of about $0.6 M_\odot$ is more than $\sim 4 M_\odot$, the lowest luminosity degenerate in all clusters is of the same

Figure 2: Colour Magnitude data for NGC 2420. The data are in the WFPC1 ground photometric system. The cluster reddening vector (solid line) and the data completeness limits for V (dots) and I (dashes) are shown. We place the white dwarf detection limit near V=25, V−I=0.8

TABLE 1: Observational and Derived Results

| Cluster | $\ell$ | $b$ | [Fe/H] | $E_{B-V}$ | m–M | $\tau_{\text{isoch}}$ (Gyr) | $ref$ | $M_{\text{bol, WD}}$ (Predicted) | $M_{\text{bol, WD}}$ (Mean of) (faintest WDs) | $M_{\text{bol, WD}}$ (Coolest) (single WD) |
|---|---|---|---|---|---|---|---|---|---|---|
| NGC2477 | 254° | −6° | $\sim 0$ | $\approx 0.30$ | 10.6 | $0.6 \pm 0.1$ | 1 | 11.5 | $13.1 \pm 0.1$ | $13.1 \pm 0.1$ |
|  |  |  |  |  |  | $1.2 \pm 0.3$ | 2 | 12.9 | $13.1 \pm 0.1$ | $13.1 \pm 0.1$ |
| NGC2420 | 198° | +20° | $-0.42 \pm 0.07$ | 0.05 | 11.95 | 2.1 | 1 | 13.6 | $12.9 \pm 0.1$ | $13.1 \pm 0.1$ |
|  |  |  |  |  |  | $3.4 \pm 0.6$ | 3 | 14.4 | $12.9 \pm 0.1$ | $13.1 \pm 0.1$ |
|  |  |  |  |  |  | 4 | 4 | 14.6 | $12.9 \pm 0.1$ | $13.1 \pm 0.1$ |

REFS: (1) Carraro & Chiosi 1994; (2) Smith & Hesser 1983; (3) Anthony Twarog *et al* 1990; (4) Vandenberg 1985.

mass, $0.6M_\odot$. Only in the case of the shortest evolutionary time scale for NGC2477 (0.6Gyr, from Carraro & Chiosi) is the pre-white dwarf evolutionary time a significant fraction of the total cluster age.

Calculation of the bolometric luminosity of the coolest white dwarf in the two clusters for all the derived evolutionary ages listed in Table 1 is then straightforward. One calculates the difference in pre-white dwarf evolutionary time as a function of initial stellar mass relative to a fiducial $8M_\odot$ initial mass progenitor generating a $1M_\odot$ white dwarf. Applying a simple initial-mass – final-mass relation, one then calculates the resulting white dwarf luminosity after a time equal to the cluster isochrone age minus the additional pre-white dwarf evolutionary time. This is done for all relevant initial masses, thus identifying both the mass (in all cases near $0.6M_\odot$) and the bolometric luminosity of the end of the degenerate sequence. This value is listed in the last column of table 1.

The initial mass–final mass relation adopted here, following Weidemann (1990) assumes a final mass of $0.6M_\odot$ for inital masses of up to $4.0M_\odot$, with a linear rise to a final mass of $1.0M_\odot$ for an initial mass of $8M_\odot$. In fact, the rate of cooling of a white dwarf increases so rapidly with decreasing mass, that all models of relevance here have the lowest luminosity degenerate at $0.6M_\odot$, in spite of the additional stellar evolutionary lifetime of a $4M_\odot$ predecessor. Thus, the results here are not sensitive to the details of the adopted initial-mass – final-mass relationship.

Figure 3: Mass dependent cooling tracks, from Wood 1994. The horizontal solid line indicates the observed white dwarf cooling sequence limit in the two clusters. The indicated cooling curves, from left to right at the position of the solid line, are for masses of $0.5M_\odot, 0.6M_\odot, 0.7M_\odot, 0.8M_\odot, \& 1.0M_\odot$.

### 5.1 NGC2477

There are two isochrone ages available for this cluster, a recent calculation of 0.6Gyr and an earlier calculation of 1.2Gyr. The corresponding predicted limiting white dwarf luminosities are $M_{bol} = 11.5$ and 12.9. the observed limit is marginally consistent with the latter, but excludes the former age. If the data reach the true cooling limit of the white dwarf sequence, and are not affected by incompleteness, then the age calibration of isochrone ages near 1Gyr provided by the 1977 Ciardullo & Demarque isochrones is supported. The 1994 Bertelli *etal* isochrone calibration as applied by Carraro & Chiosi is not supported by these data. We emphasise however that our present data provide only an *upper limit* to the end of the cooling sequence, so that it is also unclear if there is any support for the 1.2Gyr chronology.

### 5.2 NGC2420

Isochrone ages for this cluster are available from the VandenBerg (1985) isochrones, with and without recalibration, both in agreement that the age is near 3.5Gyr. An age of 2.1Gyr is provided by Bertelli *et al* (1994). The white dwarf limiting magnitudes for these ages are $M_{bol} = 14.6$ and 13.6 respectively. The present data are limited to 70 percent completeness at $M_{bol} = 13.2$ and 20 percent completeness at $M_{bol} = 13.5$ (see figure 2), with the rapid loss of completeness being due to the problem with 'hot' pixels. The observed sequence limit is $M_{bol} = 13.1$. An increasing luminosity function to $M_{bol} \gg 13.1$ should have provided a detectable, though incomplete, signal, which provides difficulties for both chronologies. The problems of selective incompleteness at the colours of the faintest white dwarfs because of WFPC2 hot pixels (§2) however complicate reliable simulations of the incompleteness. While our adopted incompleteness limits are believed conservative, this potential uncertainty may weaken this conclusion. Future WFPC2 data, using observational strategies optimised to obviate the hot pixel and completeness problems (eg dithering), can resolve this directly.

## 6 CONCLUSIONS

Determination of the lower limit of the white dwarf luminosity function is potentially the most precise age dating method applicable to star clusters. This method is also accurate, in that it is nearly independant of stellar evolutionary theory, and depends primarily on relatively well-understood physics, at least for ages between $\sim$ 1Gyr and $\sim$ 8Gyr. We have obtained deep HST imaging of two nearby intermediate age open clusters, have detected the white dwarf sequence unambiguously in each cluster, and have derived a lower limit on its age. These limits are not consistent with the Carraro & Chiosi (1994) isochrone age for NGC2477, and present difficulties for other isochrone calibrations.

## ACKNOWLEDGEMENTS

We thank Nial Tanvir and David Robinson of the UK HST Support Facility, Becky Elson and Basilio Santiago for advice and assistance. This *Letter* is based on observations with the NASA/ESA *Hubble Space Telescope*, obtained at the Space Telescope Science Institute, which is operated by the Association of Universities for Research in Astronomy, Inc, under NASA contract NAS5-26555.

## REFERENCES

Andersen, J. 1991 Astron Astrophys Review 3 p91
Anthony Twarog, B.J., Kaluzny, J., Shara, M.M., & Twarog, B.A. 1989 AJ 99, 1504
Bertelli, G., Bressan, A., Chiosi, C., Fagotto, F., & Nasi, E. 1994 A & A S in press
Canal, R. 1990 in *Baryonic Dark Matter* eds D. Lynden-Bell & G. Gilmore. (Kluwer, Dordrecht) p103.
Carraro, G., & Chiosi, C. 1994 A & A in press
Chieffi, A., & Straniero, O., 1989 ApJS 71, 47
Ciardullo, R.B., & Demarque, P. 1977 Trans Yale Univ Obs vol 33
Demarque, P., Green, E.M., & Guenther, D.B., 1992 AJ 103, 151
Harris, H.C., Hunter, D.A., Baum, W.A., & Jones, J.H. 1993 AJ 105, 1196
Hartwick, F.D.A., Hesser, J.E., & McClure, R.D. 1972 ApJ 174, 557
Holtzman, J.A., Hester, J.J., Casertano, S., Trauger, J.T., Bellester, G.E., Burrows, C.J., Clarke, J.T., Crisp, D., Gallagher, J.S., Griffiths, R.E., Hoessel, J.G., Mould, J.R., Scowen, P.A., Staplefeldt, K.R., Watson, A.M., & Westphal, J.A. 1995 PASP in press.
Liebert, J., Dahn, C.C., and Monet, D.G. 1988 ApJ 332, 891
Maeder, A., & Meynet, G. 1992 A& AS 89, 451
Reimers, D., & Koester, D 1982, A&A 116, 341
Smith, H.A., & Hesser, J. 1983 PASP 95, 277
VandenBerg, D.A., 1985 ApJS, 58, 711
von Hippel, T., Gilmore, G., Jones, D.H.P., Tanvir, N., & Robinson, D., 1995 in preparation.
Wood, M. 1994 in *IAU Coll 147: The Equation of State in Astrophysics* eds G. Chabrier & E. Schatzmann (CUP:

Cambridge) p612
Weidemann, V., 1990 ARAA 28, p103
Weidemann, V., & Koester, D. 1983, A&A 132, 195
Winget, D.E., Hansen, C.J., Liebert, J., van Horn, H.M.,
    Fontaine, G., Nather, R., Kepler, S.O., & Lamb, D.Q. 1987,
    ApJL 315, L77